\documentclass[11pt,a4paper, dvipsnames]{article}
\usepackage[hyperref]{emnlp-ijcnlp-2019}
\usepackage{times}
\usepackage{latexsym}
\usepackage{bbm}
\usepackage{xcolor}
\usepackage{multirow}
\usepackage{url}
\usepackage{amsmath}
\usepackage{amsfonts,amssymb}
\usepackage{graphicx}
\usepackage{xspace}
\usepackage{booktabs}

\usepackage{subcaption}
\usepackage{enumitem}
\usepackage{wrapfig,lipsum}

\setlist[itemize]{leftmargin=*}  

\newcommand{\MORES}{\textsc{mores}\xspace}
\aclfinalcopy

\title{Modularized Transfomer-based Ranking Framework}

\author{Luyu Gao \qquad 
Zhuyun Dai \qquad 
 Jamie Callan\\
  Language Technologies Institute\\
  Carnegie Mellon University \\
  {\tt \{luyug, zhuyund, callan\}@cs.cmu.edu} \\
  }

\date{}

\begin{document}
\maketitle



\begin{abstract}

Recent innovations in Transformer-based ranking models have advanced the state-of-the-art in information retrieval. However, these Transformers are computationally expensive, and their opaque hidden states make it hard to understand the ranking process. In this work, we modularize the Transformer ranker into separate modules for text \emph{representation} and \emph{interaction}. We show how this design enables substantially faster ranking using offline pre-computed representations and light-weight online interactions. The modular design is also easier to interpret and sheds light on 
the ranking process in Transformer rankers.\footnote{Open source code at \url{https://github.com/luyug/MORES}} 
\end{abstract}

%


\maketitle

\section{Introduction}
\label{sec-introduction}



Neural rankers based on Transformer architectures \cite{Vaswani2017AttentionIA} fine-tuned from BERT \cite{Devlin2019BERTPO} achieve current state-of-the-art~(SOTA) ranking effectiveness \cite{nogueira2019passage,craswell2019overview}. The power of the Transformer comes from self-attention, the process by which all possible pairs of input tokens interact to understand their connections and contextualize their representations. Self-attention provides detailed, token-level information for matching, which is critical to the effectiveness of Transformer-based rankers~\cite{wu2019investigating}. 

When used for ranking, a Transformer ranker takes in the concatenation of a query and document, applies a series of self-attention operations, and outputs from its last layer a relevance prediction~\cite{nogueira2019passage}. The entire ranker runs like a black box and hidden states have no explicit meanings. This represents a clear distinction from earlier neural ranking models that keep separate text representation and distance~(interaction) functions.  Transformer rankers are slow~\cite{nogueira2019doc2qry}, and the black-box design makes it hard to interpret their behavior.

We hypothesize that a Transformer-based ranker simultaneously performs text representation and query-document interaction as it processes the concatenated pair. Guided by this hypothesis, we decouple representation and interaction with a \emph{MO}dualarized  \emph{RE}ranking \emph{S}ystem (\MORES). \MORES consists of three \emph{Transformer} modules: the Document Representation Module, the Query Representation Module, and the Interaction Module. The two Representation Modules run independently of each other. The \textbf{Document Representation Module} uses self-attention to embed each document token conditioned on all document tokens.  The \textbf{Query Representation Module} embeds each query token conditioned on all query tokens. The \textbf{Interaction Module} performs attention from \emph{query} representations to \emph{document} representations to generate match signals and aggregates them through self-attention over query tokens to make a relevance prediction.

By disentangling the Transformer into modules for representation and interaction, \MORES can take advantage of the indexing process: while the interaction must be done online, document representations can be computed offline. We further propose two strategies to pre-compute document representations that can be used by the Interaction Module for ranking.

Our experiments on a large supervised ranking dataset demonstrate the effectiveness and efficiency of \MORES.  It is as effective as a state-of-the-art BERT ranker and can be up to $120\times$ faster at ranking.  A domain adaptation experiment shows that the modular design does not affect the model transfer capability, so \MORES can be used under low-resource settings with simple adaptation techniques. 
By adapting individual modules, we discovered differences between representations and interaction in adaptation. The modular design also makes \MORES more interpretable, as shown by our attention analysis, providing new understanding of black-box Transformer rankers.
%

\section{Related Work}
\label{sec-background}

Neural ranking models for IR proposed in previous studies can be generally
classified into two groups~\cite{DRMM}: \emph{representation-based} models, and \emph{interaction-based} models. 

\emph{Representation-based} models learn latent vectors (embeddings)
of queries and documents and
use a simple scoring function (e.g., cosine) to measure the relevance
between them.  Such methods date back to LSI~\cite{deerwester1990indexing} and classical siamese networks~\cite{bromley1993siamese}.  More recent research considered using modern deep learning techniques to learn the representations. Examples include DSSM~\cite{huang2013learning}, C-DSSM~\citep{shen2014learning}, etc.  Representations-based models are efficient during evaluation because the document representations are independent of the query, and therefore can be pre-computed. However, compressing a document into a single low-dimensional vector loses specific term matching signals~\citep{DRMM}. As a result, previous representation-based ranking models mostly fail to outperform interaction-based ones.  

\emph{Interaction-based} models, on the other hand, use a neural network to model the word-level interactions between the query and the document. Examples include  DRMM~\cite{DRMM} and K-NRM~\cite{K-NRM}. Recently, Transformers~\cite{Vaswani2017AttentionIA}, especially BERT~\cite{Devlin2019BERTPO} based Transformers, have been widely used in information retrieval ranking tasks~\cite{nogueira2019passage, dai2019deeper, qiao2019understanding}. BERT-based rankers concatenate query and document into a single string and apply self-attention that spans over the query and the document in every layer. 
Rankers using pre-trained Transformers such as BERT has become the current state-of-the-art~\cite{craswell2019overview}. However, the performance gains come at the computational cost of inferring the many token-level interaction signals at the evaluation time, which scales quadratically to the input length.
It is an open question whether we can \emph{combine} the advantages of representation-based and interaction-based approaches. Little research has studied this direction prior to this work.

There are several research directions aiming to reduce the computational cost of Transformer models. One line of research seeks to compress the big Transformer into smaller ones using model pruning~\cite{voita2019headpruning} or knowledge distillation~\cite{Hinton2015DistillingTK, Sanh2019DistilBERTAD}.  Another line of research aims to develop new Transformer-like units that have lower complexity than the original Transformer. For example, ~\cite{child2019sparsetransformer} introduces sparse factorizations of the attention matrix which efficiently compute subsets of the attention matrix. The focus of this work is an efficient framework to combine Transformers for ranking; all aforementioned techniques can be applied to individual Transformers within our framework, and are therefore orthogonal to this paper.



\section{Proposed Method}
\label{sec-method}




In this section, we introduce the Modularized Reranking System~(\MORES),  how \MORES can speed up retrieval, 
and how to effectively train and initialize \MORES.

\subsection{The MORES Framework}
\label{subsec: earl}

A typical Transformer ranker takes in the \emph{concatenation} of a query $qry$ and a document $doc$ as input. 
At each layer, the Transformer generates a new contextualized embedding for each token based on its attention to all tokens in the concatenated text. This formulation poses two challenges. First, in terms of speed, the attention consumes time quadratic to the input length. As shown in Table~\ref{tab:complexity}, for a query of $q$ tokens and a document of $d$ tokens, the Transformer would require assessments of $(d+q)^2$ pairs of tokens. Second, as query and document attention is entangled from the first layer, it is challenging to interpret the model.



\begin{figure}[t]
\centering
\caption{An illustration of the attention within a \MORES model using two layers of Interaction Blocks ($2\times$ IB). Representation Modules only show 1 layer of attention due to space limits. In a real model, Document Representation Module and Query Representation Module are deeper than shown here.}
\includegraphics[width=0.46\textwidth]{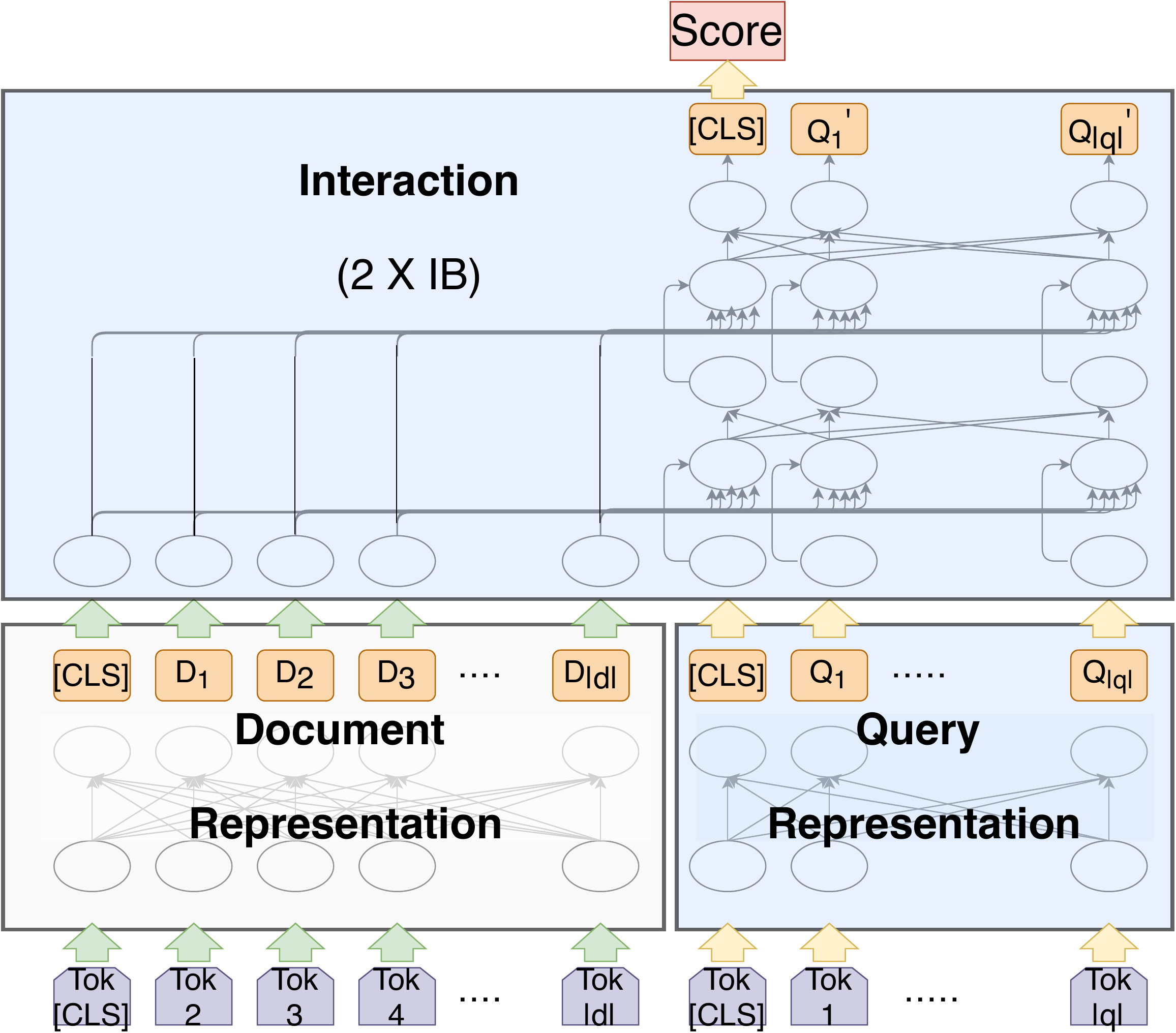}
\label{fig: attn}
\end{figure}

\MORES aims to address both problems by disentangling the Transformer ranker into document representation, query representation, and interaction, each with a dedicated Transformer, as shown in \autoref{fig: attn}. The document representation is query-agnostic and can be computed off-line. The interaction uses query-to-document attention, which further reduces online complexity. This separation also assigns roles to each module, making the model more transparent and interpretable.

\par The two \textbf{Representation Modules}
use 
Transformer encoders~\cite{Vaswani2017AttentionIA} to embed documents and queries respectively and independently. In particular, for documents,
\begin{align}
    H^{doc}_l &= \text{Encoder}^{doc}_l(H^{doc}_{l-1}) \\
    H^{doc}_1 &= \text{Encoder}^{doc}_1(\text{lookup}(doc))
\end{align}
and for queries,
\begin{align}
    H^{qry}_l &= \text{Encoder}^{qry}_l(H^q_{l-1}) \\
    H^{qry}_1 &= \text{Encoder}^{qry}_1(\text{lookup}(qry))
\end{align}
where lookup represents word\footnote{We use WordPiece tokens, following BERT.} and position embeddings, and Encoder represents a Transformer encoder layer. Query and document Representation Modules can use different numbers of layers.  Let $M$ and $N$ denote the number of layers for document and query representations respectively.  The hidden states from the last layers are used as the Representation Modules' output. Formally, for a document of length $d$, query of length $q$, and model dimension $n$, let matrix $D=H_M^{doc} \in \mathbb{R}^{d \times n}$ be the output of the Document Representation Module and $Q = H^{qry}_N \in \mathbb{R}^{q \times n}$ be the output of the Query Representation module.

\begin{table*}[t]
\caption{Time complexity of \MORES and a typical Transformer ranker, e.g., a standard BERT ranker. We write $q$ for query length, $d$ for document length, $n$ for Transformer's hidden layer dimension, and $N_{doc}$ for number of candidate documents to be ranked for each query. For interaction, Reuse-S1 corresponds to document representation reuse strategy, and Reuse-S2 projected document representation reuse strategy.
}
\centering
\small
\begin{tabular}{ l | c || c | c }
\hline \hline 
 & Total,  &Online,  & Online, \\
 & 1 Query-Document Pair & 1 Query-Document Pair  &$N_{doc}$ Documents\\
\hline
Typical Transformer Ranker & $n(d+q)^2 + n^2(d+q)$ & $n(d+q)^2 + n^2(d+q)$ & $ (n(d+q)^2 + n^2(d+q)) N_{doc}$ \\ \hline 
Document Representation & $nd^2 + n^2d$ & 0 & 0\\
Query Representation & $nq^2 + n^2q$ & $nq^2 + n^2q$ & $(nq^2 + n^2q) $ \\
Interaction w/ Reuse-S1 & $n(qd+ q^2) + n^2(q+d)$ & $n(qd+ q^2) + n^2(q+d)$ & $ (n(qd+ q^2) + n^2(q+d))  N_{doc}$\\
Interaction w/ Reuse-S2 & $n(qd+ q^2) + n^2(q+d)$ & $n(qd+ q^2) + n^2q$ & $ (n(qd+ q^2) + n^2q)  N_{doc}$\\

\hline \hline 
\end{tabular}
\label{tab:complexity}
\end{table*}

\par The \textbf{Interaction Module}  
uses the Representation Modules' outputs, $Q$ and $D$, to make a relevance judgement. The module consists of a stack of Interaction Blocks~(IB), a novel attentive block that performs query-to-document cross-attention, followed by query self-attention\footnote{We use multi-head version of attention in the Interaction Blocks (IB).}, as shown in \autoref{fig: attn}. Here, we write cross-attention from $X$ to $Y$ as $\text{Attend}(X, Y)$, self-attention over $X$ as $\text{Attend}(X, X)$ and layer norm as LN. Let,
\begin{align}
    Q_{\text{x}} &=  \text{LN}(\text{Attend}(Q, D) + Q) \label{eq:cross} \\
       Q_{\text{self}} &=  \text{LN}\left(\text{Attend}(Q_{\text{x}}, Q_{\text{x}} \right) + Q_{\text{x}}) \label{eq:self}
\end{align}
\autoref{eq:cross} models interactions from query tokens to document token. Each query token in $Q$ attends to document embeddings in $D$ to produce relevance signals. Then, \autoref{eq:self} collects and exchanges signals among query tokens by having the query tokens attending to each other. 
The output of the first Interaction Block (IB) is then computed with a feed-forward network~(FFN) on the query token embeddings with residual connections,
\begin{align}
    \text{IB}(Q, D) = \text{LN}\left(\text{FFN}\left(Q_{\text{self}}\right) + Q_{\text{self}}\right)
\end{align}
We employ multiple Interaction Blocks to iteratively repeat this process and refine the hidden query token representations, modeling multiple rounds 
of interactions, producing a series of hidden states, \emph{while keeping document representation $D$ unchanged},
\begin{align}
    H_l^{IB} &= \text{IB}_l(H^{IB}_{l-1}, D) \\
    H_1^{IB} &= \text{IB}_1(Q, D)
\end{align}
The Interaction Block (IB) is a core component of \MORES.  As shown in Table~\ref{tab:complexity}, its attention avoids the heavy full-attention over the concatenated query-document sequence, i.e.~$(d+q)^2$ terms, saving online computation. 

To induce relevance, we project the [CLS] token's embedding in the last~($K^\text{th}$) IB's output to a score,
\begin{equation}
    score(qry, doc) = \textbf{w}^T \text{CLS}(H^{\text{IB}}_K)
\end{equation}

\subsection{Pre-Compute and Reuse Representation}
\label{sec-reuse}


\MORES's modular design allows us to pre-compute and reuse representations. The Query Representation Module runs once when receiving the new query; the representation is then repeatedly used to rank the candidate documents. More importantly, the document representations can be built offline. We detail two representation reuse strategies with different time vs.~space trade-offs: 1) a document representation reuse strategy that stores the Document Representation Module's output, and 2) a projected document representation reuse strategy that stores the Interaction Module's intermediate transformed document representations. These strategies have the same overall math, produce \emph{the same} ranking results, and only differ in time/space efficiency. 

\textbf{Document Representation Reuse Strategy~(Reuse-S1)} runs the Document Representation Module offline, pre-computing document representations $D$ for all documents in the collection. 
When receiving a new query, \MORES looks up document representations $D$ for candidate documents, runs the Query Representation Module to get a query's representation $Q$, and feeds both to the Interaction Module to score. This strategy reduces computation by not running the Document Representation Module at query time.

\textbf{Projected Document Representation Reuse Strategy~(Reuse-S2)} further moves document-related computation performed in the Interaction Module offline. In an IB, the cross-attention operation first projects document representation $D$ with key and value linear projections~\cite{Vaswani2017AttentionIA}
\begin{align}
   D_k =  DW_k ,\; D_v = DW_v \label{eq:s2}
\end{align}
where $W_k , W_v$ are the projection matrices.
For each IB, Reuse-S2 pre-computes and stores $D_{proj}$\footnote{We pre-compute for all attention heads in our multi-head implementation},
\begin{equation}
    D_{proj} = \{DW_k , DW_v \}
\end{equation}
Using Reuse-S2, 
the Interaction Module no longer needs to compute the document projections at online evaluation time. Reuse-S2 takes more storage: for each IB, both key and value projections of $D$ are stored, meaning that an Interaction Module with $l$ IBs will store $2l$ projected versions of $D$.    
With this extra pre-computation, Reuse-S2 trades storage for further speed-up.

\autoref{tab:complexity} analyzes the online time complexity of \MORES and compares it to the time complexity of a standard BERT ranker. 
We note that \MORES can move all document only computation offline. Reuse-S1 avoids the document self attention term $d^2$, which is often the most expensive part due to long document length. 
Reuse-S2 further removes from online computation the document transformation term $n^2d$, one that is linear in document length and quadratic in model dimension. 

\subsection{MORES Training and Initialization}
\label{subsec: training-techniques}

\MORES needs to learn three Transformers: two Representation Modules and one Interaction Module. The three Transformer modules are \emph{coupled} during training and \emph{decoupled} when used. To train \MORES, we connect the three Transformers and enforce module coupling with end-to-end training using the pointwise loss function~\cite{dai2019deeper}. When training is finished, we store the three Transformer modules separately and apply each module at the desired offline/online time. 

We would like to use pre-trained LM weights to ease optimization and improve generalization. However, there is no existing pre-trained LM that involves cross-attention interaction that can be used to initialize the Interaction Module. To avoid expensive pre-training, we introduce BERT weight assisted initialization. We use one copy of BERT weights to initialize the Document Representation Module. We split another copy of BERT weights between Query Representation and Interaction Modules. For \MORES with $l$ IBs, the first $12 - l$ layers of the BERT weights initialize the Query Representation Module, and the remaining $l$ layers' weights initialize the Interaction Module. This initialization scheme ensures that Query Representation Module and the IBs use consecutive layers from BERT. As a result, upon initialization, the output of the Query Representation Module and the input of the first IB will live in the same space. In addition, for IBs, query to document attention initializes with the same BERT attention weights as query self-attention. In practice, we found initializing query to document attention weights important; random initialization leads to substantially worse performance. Details can be found in \autoref{sec:rank-effectiveness}.

\section{Effectiveness and Efficiency in Supervised Ranking}
\label{sec:eff2}
The first experiment compares the effectiveness and efficiency of \MORES to a state-of-the-art BERT ranker for supervised ranking. 

\begin{table*}[t]
\caption{Effectiveness of \MORES models and baseline rankers on the MS MARCO Passage Corpus. $*$ and $\dagger$ indicate non-inferiority (Section \ref{subsec: main-setup}) with $p<0.05$ to the BERT ranker using a 5\% or 2\% margin, respectively.}
\centering
\renewcommand{\arraystretch}{0.9}

\begin{tabular}{ l || c | c c c}
\hline \hline
 & \multicolumn{4}{c}{MS MARCO Passage Ranking}\\ 
\hline
  & Dev Queries & \multicolumn{3}{c}{ TREC2019 DL Queries}  \\ 
 
Model & MRR & MRR & NDCG@10 & MAP \\  
 \hline
 BERT ranker  & 0.3527  & 0.9349 & 0.7032 & 0.4836  \\
 \hline
\MORES $1\times$ IB & $0.3334^*$ & $0.8953^*$	& $0.6721^*$ & $0.4516^*$ \\
\MORES $2\times$ IB & $0.3456^\dagger$ & $0.9283^\dagger$ & $0.7026^\dagger$ & $0.4777^\dagger$ \\
\MORES $3\times$ IB & $0.3423^\dagger$ & $0.9271^\dagger$ & $0.6980^\dagger$ & $0.4687^*$ \\
\MORES $4\times$ IB & $0.3307^*$ & $0.9322^\dagger$ & $0.6565^*$ & $0.4559^*$ \\
 \hline \hline 
\end{tabular}
\label{tab:perf-marco}
\end{table*}

\subsection{Setup}
\label{subsec: main-setup}

We use the \textbf{MS MARCO passage ranking collection} (MS MARCO)
~\cite{nguyen2016ms} 
and evaluate on two query sets with distinct characteristics:
\emph{Dev Queries} have a single relevant document with a binary relevance label. Following~\citet{nguyen2016ms}, we used MRR@10 to evaluate the ranking accuracy on this query set. \emph{TREC2019 DL Queries} is the evaluation set used in the TREC 2019 Deep Learning Track.
Its queries have multiple relevant documents with graded relevance.
Following~\citet{craswell2019overview}, we used MRR, NDCG@10, and MAP@1000 as evaluation metrics. All methods were evaluated in a \emph{reranking} task to re-rank the top 1000 documents of the MS MARCO official BM25 retrieval results.

We test \MORES effectiveness with a varied number of Interaction Blocks (IB) to study the effects of varying the complexity of query-document interaction. Models using 1 layer of IB~($1\times$ IB) up to 4 layers of IB ~($4\times$ IB) are tested. 

We compare \MORES with the BERT ranker, a state-of-the-art ranker fine-tuned from BERT, which processes concatenated query-document pairs. Both rankers are trained with the MS MARCO training set consisting of single relevance queries. 
We train \MORES on a 2M subset of Marco's training set. We use stochastic gradient descent to train the model with a batch size of 128. We use AdamW optimizer with a learning rate of 3e-5, a warm-up of 1000 steps and a linear learning rate scheduler for all \MORES variants. Our baseline BERT model is trained with similar training setup to match performance reported by \citet{nogueira2019passage}.
Our BERT ranker re-implementation has better performance compared to that reported by \citet{nogueira2019passage}.  The BERT ranker and all \MORES models are implemented with Pytorch \cite{pytorch} based on the huggingface implementation of Transformers \cite{Wolf2019HuggingFacesTS}.

We aim to test that \MORES' accuracy is equivalent to the original BERT ranker (while achieving higher efficiency). To establish equivalence,
statistical significance testing was performed with a non-inferiority test commonly used in the medical field to test that two treatments have similar effectiveness~\cite{jayasinghe2015statistical}. In this test, rather than testing to reject the null hypothesis $H_0$: $\mu_{\text{BERT}}=\mu_{\MORES}$, we test to reject $H_0'$
: $\mu_{\text{BERT}} - \mu_{\MORES} > \delta$ for some small margin $\delta$. By rejecting $H_0'$ we accept the alternative hypothesis, which is that
any reduction of performance in \MORES compared to the original BERT ranker is inconsequential. We set the margin $\delta$ to 2\% and 5\% of the mean of the BERT ranker.

\subsection{Ranking Effectiveness} 
\label{sec:rank-effectiveness}

\autoref{tab:perf-marco} reports the accuracy of \MORES and the baseline BERT-based ranker. %
 The experiments show that \MORES with $1\times$ IB can achieve $95\%$ of BERT performance. \MORES with $2\times$ IB can achieve performance comparable to the BERT ranker with a $2\%$ margin.  Three IBs does not improve accuracy and four hurts accuracy. We believe that this is due to increased optimization difficulties which outweighs improved model capacity.  Recall that for \MORES we have one set of artificial cross attention weights \emph{per IB} not initialized with real pre-trained weights. 
Performance results are consistent across the two query sets, showing that \MORES can identify strong relevant documents (Dev Queries), and can also generalize to ranking multiple, weaker relevant documents (TREC2019 DL Queries).

The results show that \MORES can achieve ranking accuracy competitive with state-of-the-art ranking models, and suggest
that the entangled and computationally expensive full-attention Transformer can be replaced by \MORES's lightweight, modularized design.  Document and query representations can be computed independently without seeing each other. With the contextualized representation, 2 layers of lightweight interaction are sufficient to estimate relevance. 

We also investigate IB initialization and compare \MORES $2\times$ IB initialized by our proposed initialization method~(copy self attention weight of BERT as IB cross attention weight), with a random initialization method~(cross attention weights randomly initialized). \autoref{tab: weight-init} shows that random initialization leads to a substantial drop in performance, likely due to difficulty in optimization.

\begin{table}
\caption{Ranking Accuracy of \MORES when using / not using attention weights copied from BERT to initialize Interaction Module. The models were tested on the MS MARCO dataset with the Dev Queries.  }
\centering
\scalebox{0.8}{
\begin{tabular}{ l | c | c c c }
\hline 
 & Dev Queries & \multicolumn{3}{c}{TREC2019 DL} \\

 & MRR@10 & MRR & NDCG@10 & MAP \\
 \hline
copy  & 0.3456 & 0.9283 & 0.7026 & 0.4777\\
random  & 0.2723 & 0.8430 & 0.6059 & 0.3702\\
 \hline
\end{tabular}
}
\label{tab: weight-init}
\end{table}

\subsection{Ranking Efficiency}

\begin{table}[t]

\caption{Average time in seconds to evaluate one query with 1,000 candidate documents, and the space used to store pre-computed representations for each document. Len: input document length. }
\label{tab:speed}
\subcaption{Document Representation Reuse (Reuse-S1)}
\resizebox{0.49\textwidth}{!}{
\begin{tabular}{ c | l || r r | r r | r }
\hline \hline 
 & & \multicolumn{2}{c|}{CPU} & \multicolumn{2}{c|}{GPU} & Space\\ 
 Len & Model  & \multicolumn{2}{c|}{Time }  & \multicolumn{2}{c|}{Time } &  (MB)  \\ 
 \hline
  \multirow{3}{*}{128} & BERT ranker  & 161s & - &2.70s & - & 0\\
&  \MORES $1\times$IB &  4s & 40x & 0.04s & 61x & 0.4 \\
  &\MORES $2\times$IB & 8s & 20x & 0.12s & 22 x & 0.4 \\
\hline
 \multirow{3}{*}{512} & BERT ranker &  698s & - & 13.05s & - & 0\\
 &\MORES $1\times$IB &  11s & 66x & 0.14s & 91x & 1.5\\
 &\MORES $2\times$IB  &  20s & 35x & 0.32s & 40x & 1.5\\
 \hline \hline
\end{tabular}}

\subcaption{ Projected Document Representation Reuse (Reuse-S2)}

\resizebox{0.49\textwidth}{!}{
\begin{tabular}{ c | l || r r | r r | r }
\hline \hline 
 & & \multicolumn{2}{c|}{CPU} & \multicolumn{2}{c|}{GPU} & Space\\ 
 Len & Model  & \multicolumn{2}{c|}{Time }  & \multicolumn{2}{c|}{Time } &  (MB)  \\ 
 \hline
  \multirow{3}{*}{128} & BERT ranker & 161s & - & 2.70s & - & 0\\
  &\MORES $1\times$IB & 2s & 85x & 0.02s & 118x & 1.5\\
  &\MORES $2\times$IB & 5s & 36x & 0.05s & 48x & 3.0 \\
  \hline
  \multirow{3}{*}{512} & BERT ranker &698s & - & 13.05s & - & 0 \\
   &\MORES $1\times$IB & 3s & 170x & 0.08s & 158x & 6.0 \\
   &\MORES $2\times$IB & 6s & 124x & 0.10s & 124x & 12.0 \\
 \hline \hline 
\end{tabular}}
\end{table}

\begin{table*}[t]

\caption{Domain adaptation on ClueWeb09-B. adapt-interaction and adapt-representation use \MORES $2\times$ IB. $*$ and $\dagger$ indicate non-inferiority (Section \ref{subsec: main-setup}) with $p<0.05$ to the BERT ranker using a 5\% or 2\% margin, respectively.}
\centering
\renewcommand{\arraystretch}{0.9}

\begin{tabular}{ l || c c c || c c c }
\hline \hline
    & \multicolumn{6}{c}{Clueweb09-B} \\ \cline{2-7}
    & \multicolumn{3}{c||}{Title Queries} & \multicolumn{3}{c}{Description Queries} \\ \cline{2-7}
 & NDCG@20 & MAP & Prec@20 & NDCG@20 & MAP & Prec@20  \\  
 \hline
 BERT ranker  & $0.3294$  & $0.1882$ & $0.3755$ & $0.3597$ & $0.2075$ & $0.3881$ \\
 \hline
 \MORES $1\times$ IB & $0.3059$ & $0.1753$ & $0.3407$ & $0.3472$ & $0.2009$ & $0.3705$\\
 \MORES $2\times$ IB & $0.3317^\dagger$ & $0.1872^\dagger$ & $0.3662^\dagger$ & $0.3571^\dagger$ & $0.2039^\dagger$ & $0.3816^\dagger$\\
 \MORES $3\times$ IB & $0.3299^\dagger$ & $0.1841^\dagger$ & $0.3679^\dagger$ & $0.3476^*$ & $0.2008^*$ & $0.3763^*$\\
 \MORES $4\times$ IB & $0.3164^*$ & $0.1824^*$ & $0.3515$ & $0.3472^*$ & $0.2012^*$ & $0.372^*$\\
 \hline
 adapt-interaction & $0.3179^*$ & $0.1849^\dagger$ & $0.3548$ & $0.3385$ & $0.1976^*$ & $0.3652$\\
 adapt-representation & $0.3319^\dagger$ & $0.1865^\dagger$ & $0.3657^*$ & $0.3557^\dagger$ & $0.2072^\dagger$ & $0.3828^\dagger$\\
 \hline \hline 
\end{tabular}
\label{tab:perf-cw}
\end{table*}

\begin{table*}[t]
\caption{Domain adaptation on Robust04. adapt-interaction and adapt-representation use \MORES $2\times$ IB. $*$ and $\dagger$ indicate non-inferiority (Section \ref{subsec: main-setup}) with $p<0.05$ to the BERT ranker using a 5\% or 2\% margin, respectively.}
\centering
\renewcommand{\arraystretch}{0.9}
\begin{tabular}{ l || c c c || c c c }
\hline \hline
    & \multicolumn{6}{c}{Robust04} \\ \cline{2-7}
    & \multicolumn{3}{c||}{Title Queries} & \multicolumn{3}{c}{Description Queries} \\ \cline{2-7}
 & NDCG@20 & MAP & Prec@20 & NDCG@20 & MAP & Prec@20  \\  
 \hline
 BERT ranker  & 0.4632  & 0.2225 & 0.3958 & 0.5065 & 0.245 & 0.4147 \\
 \hline
 \MORES $1\times$ IB & $0.4394^*$ & $0.2097$ & $0.3741^*$ & $0.4683$ & $0.2263$ & $0.3835$ \\
 \MORES $2\times$ IB & $0.4599^\dagger$  & $0.2194^\dagger$ & $0.3940^\dagger$ & $0.4846^*$ & $0.2323^*$ & $0.4008^*$ \\
 \MORES $3\times$ IB & $0.4551^\dagger$	& $0.2135^*$ & $0.3934^\dagger$ & $0.4854^*$ & $0.2334^*$ & $0.4006^*$ \\
 \MORES $4\times$ IB & $0.4553^\dagger$  & $0.2177^\dagger$ & $0.3938^\dagger$ & $0.4802$ & $0.2309$ & $0.3980^*$ \\
 \hline
 adapt-interaction    & $0.4389$ & $0.2117^*$ & $0.3723$ & $0.4697$ & $0.2249$ & $0.3896$\\
 adapt-representation & $0.4564^\dagger$ & $0.2182^\dagger$ & $0.3926^\dagger$ & $0.4884^*$ & $0.2327^*$ & $0.4042^*$\\
 \hline \hline 
\end{tabular}
\label{tab:perf-robust}
\end{table*}

Section~\ref{sec-reuse} introduces two representation reuse strategies for \MORES with different time vs. space trade-offs. This experiment measures \MORES' real-time processing speeds with these two strategies and compares them with measurement for the BERT ranker. We test \MORES $1\times$ IB and \MORES $2\times$ IB.  Additional IB layers incur more computation but do not improve effectiveness, and are hence not considered.
We record average time for ranking one query with 1000 candidate documents on an 8-core CPU and a single GPU.\footnote{Details are in Appendix \ref{sec:appendix-eff}.}
We measured ranking speed with documents of length 128 and 512 with a fixed query length of 16. Tables~\ref{tab:speed} (a) and (b) show the speed tests for the two reuse strategies, respectively. We also include per document data storage size~\footnote{We report un-compressed values. Compression can further reduce data storage.}.

We observe a substantial speedup in \MORES compared to the BERT ranker, and the gain is consistent across CPUs and GPUs. The original BERT ranker took hundreds of seconds -- several minutes -- to generate results for one query on a CPU machine, which is impractical for real-time use. 
Using Reuse-S1, \MORES with $1\times$ IB was $40$x faster than the BERT ranker on shorter documents ($d=128$); the more accurate $2\times$ IB model also achieved $20$x speedup. The difference is more profound on longer documents. As the length of the document increases, a larger portion of compute in BERT ranker is devoted to performing self-attention over the document sequence. \MORES pre-computes document representations and avoids document-side self attention, yielding up to $35$x to $90$x speedup on longer documents ($d=512$). 

Reuse-S2 -- the projected document reuse strategy -- further enlarges the gain in speed, leading to up to $170$x speedup using $1\times$ IB, and $120$x speedup using $2\times$ IB. Recall that Reuse-S2 pre-computes the document projections that will be used in \MORES' Interaction Module, which is of $n^2d$ time complexity where $n$ is the model hidden dimension (details can be found in the complexity analysis in Table~\ref{tab:complexity}). In practice, $n$ is often large, e.g., our experiment used $n=768$\footnote{This follows model dimension in BERT}. Reuse-S2 avoids the expensive $n^2d$ term at evaluation time. Note that Reuse-S2 \emph{does not affect accuracy}; it trades space to save more time.

\begin{figure*}[h!]
\centering
\begin{subfigure}[t]{0.24\textwidth}
        \centering
        \includegraphics[width=0.96\textwidth]{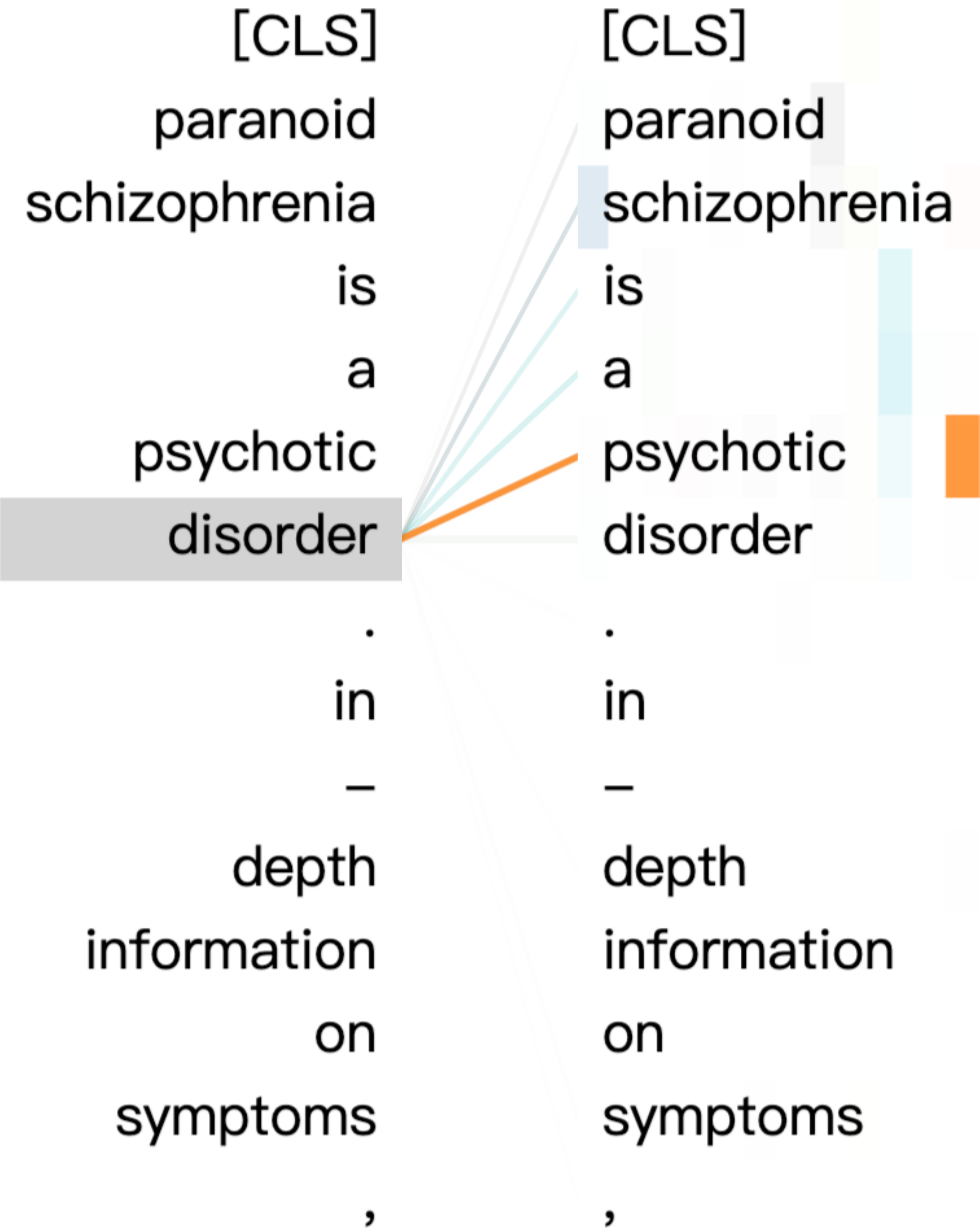}
        \caption{Document Representation}
    \end{subfigure}%
    ~
    \begin{subfigure}[t]{0.24\textwidth}
        \centering
        \includegraphics[width=0.98\textwidth]{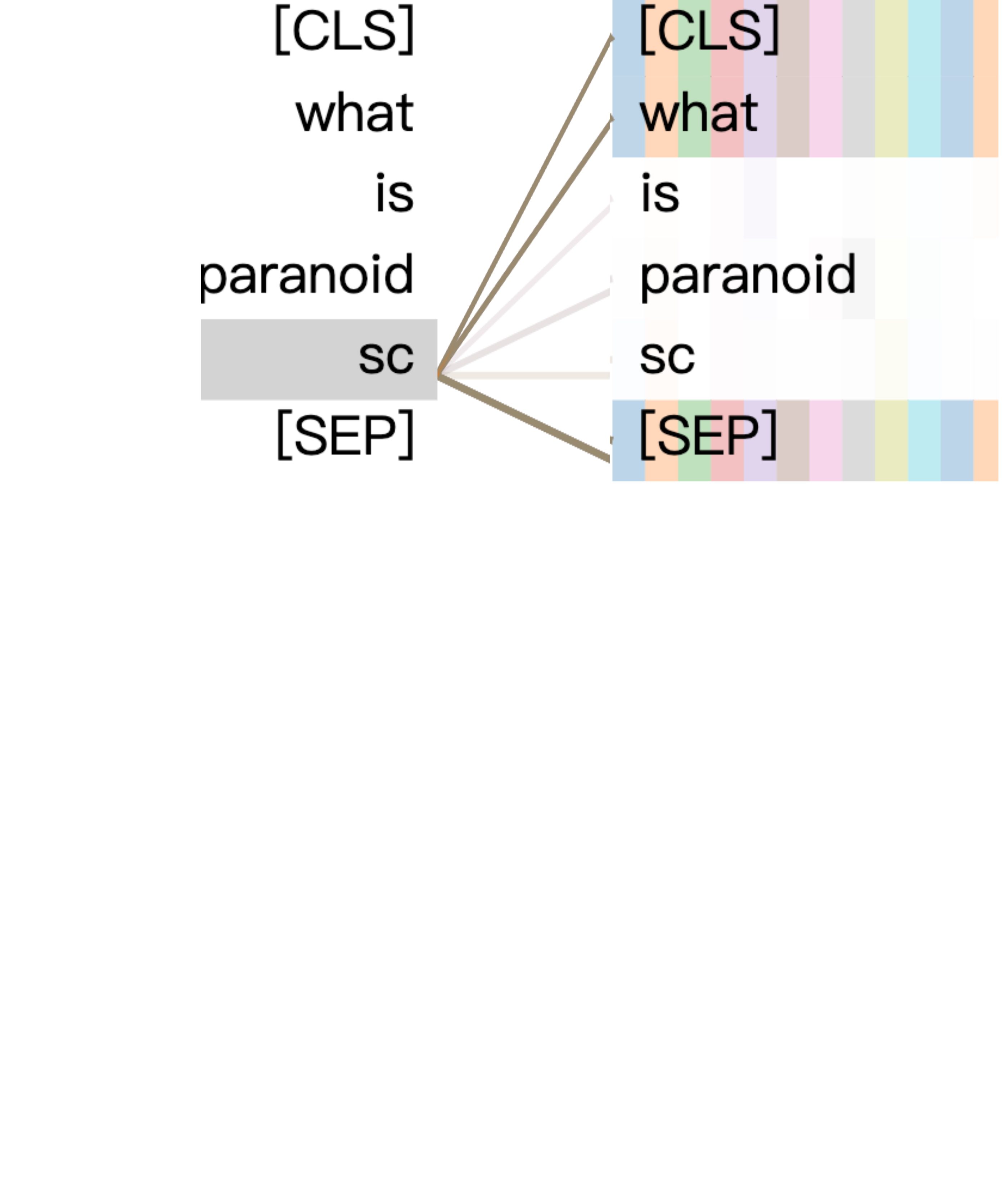}
        \caption{Query Representation}
    \end{subfigure}%
    ~
    \begin{subfigure}[t]{0.24\textwidth}
        \centering
        \includegraphics[width=0.96\textwidth]{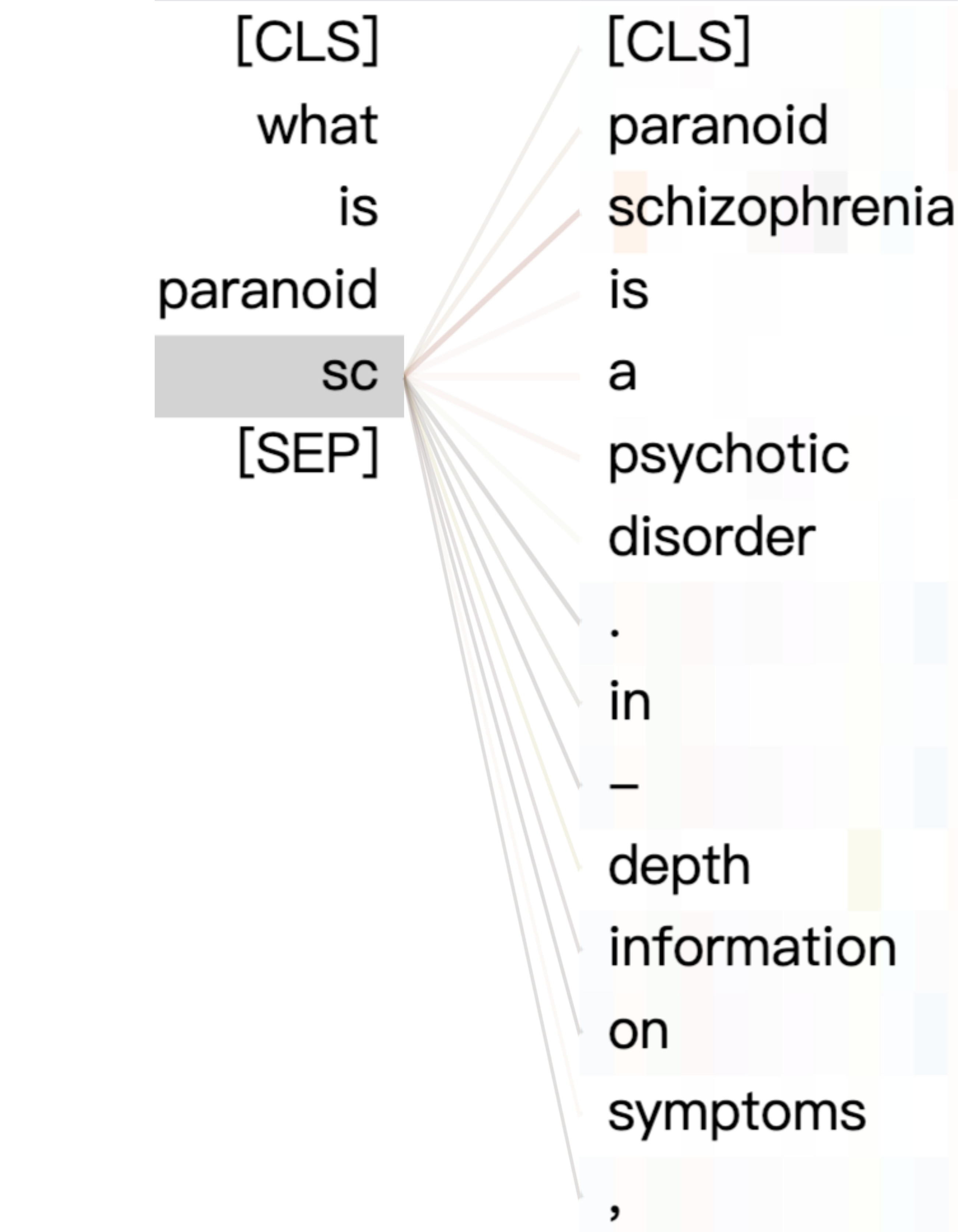}
        \caption{Interaction (1st IB)}
    \end{subfigure}%
    ~
    \begin{subfigure}[t]{0.24\textwidth}
        \centering
        \includegraphics[width=0.96\textwidth]{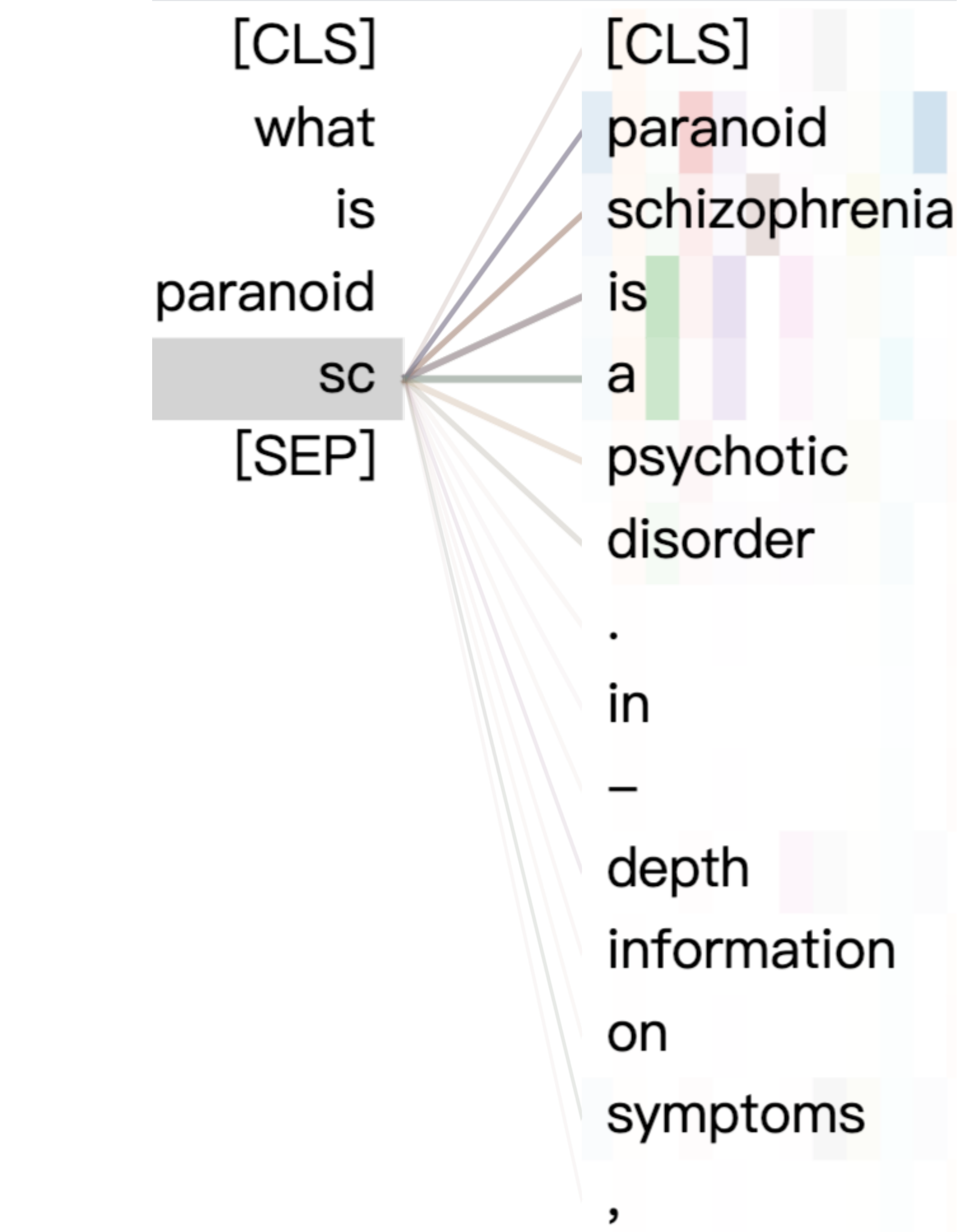}
        \caption{Interaction (2nd IB)}
    \end{subfigure}

\caption{Visualization of attention in \MORES's Representation and Interaction Modules.}
\label{fig: att-vis}
\end{figure*}

\section{Adaptation of \MORES and Modules}
The second experiment uses a domain-adaptation setting 
to investigate whether the modular design of \MORES affects adaptation and generalization ability, and how the individual Interaction and Representation Modules behave across domains.

\subsection{Setup}
This experiment trains \MORES using the MS MARCO dataset, and adapts the model to two datasets: ClueWeb09-B and Robust04. ClueWeb09-B is a standard document retrieval collection with 50M web pages crawled in 2009. Evaluation queries come from the TREC 2009-2012 Web Tracks. We used two variants of the queries: \emph{Title Queries} is 200 short, keyword-style queries. \emph{Description Queries} is 200 queries that are natural language statements or questions. Robust04 is a news corpus with 0.5M documents. Evaluation queries come from TREC 2004 Robust Track, including 250 \emph{Title Queries} and 250 \emph{Description Queries}. We evaluate ranking performance with NDCG@20, MAP, and Prec@20.

Domain adaptation is done by taking a model trained on MS MARCO and fine-tuning the model on relevant labels from the target dataset. Due to the small query sets in ClueWeb09-B and Robust04, we use 5-fold cross-validation for fine-tuning and testing. Data split, initial ranking, and document pre-processing follow~\citet{dai2019deeper}. The domain adaptation fine-tuning procedures use a batch size of 32 and a learning rate of 5e-6 while having other training settings same as supervised ranking training.

\subsection{Full Model Adaptation}
 The top 5 rows of \autoref{tab:perf-cw} and \autoref{tab:perf-robust} examine the effectiveness of adapting the full model of \MORES.  The adapted \MORES models behave similarly as on MS MARCO: using two to three layers of Interaction Blocks (IB) achieves very close to BERT ranker performance on both datasets for both types of queries while using a single layer of IB is less effective. Importantly,  our results show that the modular design of \MORES does not hurt domain transfer, indicating that new domains and low resource domains can also use \MORES through simple adaptation. 
 
 \subsection{Individual Module Adaptation}
 With separate representation and interaction components in \MORES, we are interested to see how each is affected by adaptation. We test two extra adaptation settings on \MORES $2\times$ IB:
 fine-tuning only Interaction Module on the target domain~(adapt-interaction) or only Representation Modules~(adapt-representation) on target domain. Results are shown in the bottom two rows of \autoref{tab:perf-cw} and \autoref{tab:perf-robust} for the two data sets.
 
We observe that only adapting the Interaction Module to the target domain is less effective compared to adapting the full model (\MORES $2\times$ IB), suggesting that changing the behaviour of interaction is not enough to accommodate language changes across domains. On the other hand,  freezing the Interaction Module and only fine-tuning the Representation Modules (adapt-representation) produces performance on par with full model apdatation. This result shows that it is more necessary to have domain-specific representations, while interaction patterns are more general and not totally dependent on representations.

\section{Analysis}
\label{sec:attention-analysis}


The modular design of \MORES allows Representation and Interaction to be inspected separately, providing better interpretability than a black-box Transformer ranker. 
\autoref{fig: att-vis} examines the attention with \MORES for a hard-to-understand query \textit{``what is paranoid sc"} where ``sc" is ambiguous, along with a relevant document \textit{``Paranoid schizophrenia is a psychotic disorder. In-depth information on symptoms...."}~\footnote{We only show the first 16 tokens due to space limitation.} 

In the Document Representation Module (\autoref{fig: att-vis}a), we can see that
\textit{``disorder"} uses \textit{``psychotic''} and \textit{``schizophrenia''} for contextualization, making itself more specific.  In the Query Representation Module (\autoref{fig: att-vis}b), because the query is short and lacks context, \textit{``sc''} incurs a broad but less meaningful attention.    
The query token \textit{``sc''} is further contextualized in the Interaction Module (\autoref{fig: att-vis}c) using information from the document side --  \textit{"sc"} broadly attends to the document token in the first IB to  disambiguate itself. With the extra context, \textit{``sc"} is able to correctly attend to ``schizophrenia" in the second IB to produce relevance signals (\autoref{fig: att-vis}d). 

This example explains why \MORES $1 \times$ IB performs worse than \MORES with multiple IBs -- ambiguous queries need to gather context from the document in the first IB before making relevance estimates in the second. 
More importantly, the example indicates that the query to document attention has two distinct contributions: understand query tokens with the extra context from the document, and match query tokens to document tokens, with the former less noticed in the past. We believe \MORES can be a useful tool for better interpreting and understanding SOTA black-box neural rankers.
\section{Conclusion}
\label{sec-conclusion}

State-of-the-art neural rankers based on the Transformer architecture consider all token pairs in a concatenated query and document sequence. Though effective, they are slow and challenging to interpret. 
This paper proposes \MORES, a modular Transformer ranking framework that decouples ranking into Document Representation, Query Representation, and Interaction. \MORES is effective while being efficient
and interpretable.

Experiments on a large supervised ranking task show that \MORES is as effective as a state-of-the-art BERT ranker. With our proposed document representation pre-compute and re-use methods, \MORES can achieve $120$x speedup in online ranking while retaining accuracy. 
Domain adaptation experiments show that \MORES' modular design does not hurt transfer ability, indicating that \MORES can be adapted to low-resource domains with simple techniques. 

Decoupling representation and interaction provides new understanding of
Transformer rankers. Complex full query-document attention in state-of-the-art Transformer rankers can be factored into
independent document and query representation, and shallow light-weight interaction. We further discovered two types of interaction: further query understanding based on the document, and the query to document tokens matching for relevance. Moreover, we found that the interaction in ranking is less domain-specific, while the representations need more domain adaptation. These findings provide opportunities for future work towards more efficient and interpretable neural IR.


\section*{Acknowledgments}
This work was supported in part by National Science Foundation (NSF) grant IIS-1815528. 
Any opinions, findings, and conclusions in this paper are the authors' and do not necessarily reflect those of the sponsors.
The authors would also like to thank Graham Neubig and Chenyan Xiong for helpful discussions and feedbacks.

\newpage
\bibliographystyle{acl_natbib}
\bibliography{strings-short, local}

\clearpage
\appendix
\section{Appendix}

\subsection{Implmentation Details}
\label{sec:appendix-detail}
\paragraph{Training Details}
On MS MARCO passage ranking dataset, we trained \MORES over a 2M subset of Marco's training set. We use stochastic gradient descent to train the model with a batch size of 128. We use AdamW optimizer with a learning rate of 3e-5, a warm-up of 1000 steps and a linear learning rate scheduler for all \MORES variants. Our baseline BERT model is trained with similar training setup to match performance reported in \citep{nogueira2019passage}. We have not done hyper-parameter search, and all training setup is inherited from GLUE example in the huggingface transformer code base~\cite{Wolf2019HuggingFacesTS}.
Following \cite{dai2019deeper}, we run a domain adaptation experiment on ClueWeb09-B: we take trained model on MS MARCO, and continue training over ClueWeb09-B's training data in a 5-fold cross-validation setup. We use a batch size of 32 and a learning rate of 5e-6. We select from batch size of 16 and 32, learning rate of 5e-6, 1e-5  and 2e-5 by validation point-wise accuracy.
\paragraph{Speed Test Details}
\label{sec:appendix-eff}
GPU test was run on a single RTX 2080 TI, with CUDA 10.1. We use a separate CUDA stream to pre-fetch data to the GPU. CPU tests was run in a SLURM task environment with 8 Xeon Silver 4110 logical cores.

\subsection{Parameter Details}
All \MORES models follow BERT's architecture for initialization, having 12 attention heads, 768 embedding dimension, 3072 feed forward network hidden dimension. \MORES with one IB up to four IBs have parameters of 224M, 228M, 231M and 233M parameters respectively.



\subsection{Datasets}
We use MSMARCO, ClueWeb09-b and Robust04. The first is available at \url{https://microsoft.github.io/msmarco/} and the latter two at \url{http://boston.lti.cs.cmu.edu/appendices/SIGIR2019-Zhuyun-Dai}. All input text are tokenized by BERT's WordPiece tokenizer without other pre-processing. We evaluate MS MARCO Dev query sets with its provided evaluation script and the rest with trec\_eval~(\url{https://github.com/usnistgov/trec_eval}). 

\end{document}